\documentclass[aps,pra,floatfix,twocolumn,nofootinbib,showpacs]{revtex4}

\usepackage{amsmath}
\usepackage{amssymb}
\usepackage{graphicx}
\usepackage{dcolumn}
\usepackage[sort&compress]{natbib}
\usepackage{bm}
\usepackage[dvips]{epsfig}

\usepackage{subfigure} 
\usepackage{float}

\newcommand{\comment}[1]{}
\newcommand{\BEQ}{\begin{equation}}
\newcommand{\EEQ}{\end{equation}}
\newcommand{\BEA}{\begin{eqnarray}}
\newcommand{\EEA}{\end{eqnarray}}
\newcommand{\ssum}{{\sum}}

\renewcommand{\a}{\alpha}

\begin{document}
\title{Dynamics of Boltzmann Q-Learning in \\ Two-Player Two-Action Games}

\date{\today}
\author{Ardeshir Kianercy and Aram Galstyan}
\affiliation{USC Information Sciences Institute,
Marina del Rey, CA 90292}

\begin{abstract} 
We consider the dynamics of $Q$--learning  in two--player two--action games with a Boltzmann exploration mechanism. For any non--zero exploration rate the dynamics is {\em dissipative}, which guarantees that agent strategies converge to rest points that are generally different from the game's Nash Equlibria (NE). We provide a comprehensive characterization of  the rest point structure for different games, and examine the sensitivity of this structure with respect to the noise due to exploration. Our results indicate that for a class of games with multiple NE  the asymptotic behavior of learning dynamics can undergo drastic changes at critical  exploration rates. Furthermore, we demonstrate  that for certain games with a single NE, it is possible to have  additional rest points (not corresponding to any NE) that persist  for a finite range of the exploration rates and disappear when the exploration rates  of both players tend to zero.
\end{abstract} 

\maketitle

\section{Introduction}
\noindent 

Reinforcement Learning (RL)~\cite{Sutton2000} is a powerful
framework that allows an agent to behave near--optimally  through a trial and error exploration of the
environment. Although originally developed for single agent settings, RL approaches have been extended to scenarios where multiple agents learn concurrently by interacting with each other. The  main difficulty in multi--agent learning is that, due to mutual adaptation of agents, the stationarity condition of single--agent learning environment is violated. Instead, each agent learns in a time--varying environment induced by the learning dynamics of other agents. Although in general multi--agent RL does not have any formal convergence guarantees (except in certain settings), it is known to often work well in practice.

Recently, a number of authors have addressed the issue of multi--agent learning from the perspective of dynamical systems~\cite{Claus1998,Singh2000,Bowling2001}. For instance, it has been noted that for stateless $Q$--learning with {\em Boltzmann action selection}, the dynamics of agent strategies can be described by (bi-matrix) replicator equations from population biology~\cite{Hofbauer1998}, with an additional term that accounts for the exploration~\cite{Sato2003,Sato2005,Tuyls2003}. A similar approach for analyzing  learning dynamics with $\varepsilon$-greedy exploration mechanism~\footnote{\label{foot0}The $\varepsilon$-greedy $Q$-learning schema selects the  action with highest $Q$ value with probability $(1-\epsilon)+\frac{\epsilon}{n}$ and other actions with probability of $\frac{\epsilon}{n}$, where $n$ is the number of the actions.}  was developed in~\cite{Gomes2009,Wunder2010}.

Most existing approaches so far have focused  on numerical integration or simulation methods for understanding  dynamical behavior of learning systems. Recently, ~\cite{Wunder2010} provided a full categorization of $\varepsilon$-greedy $Q$-learning dynamics in two--player two--action games using analytical insights from hybrid dynamical systems. A similar classification  for Boltzmann $Q$-learning, however, is lacking. On the other hand, a growing body of recent  neurophysiological studies indicate that  Boltzmann-type softmax action selection might be a plausible mechanism for understanding decision making in primates. For instance, experiments with monkeys playing a competitive game   indicate that  their decision making  is consistent with softmax value-based  reinforcement learning~\cite {Lee2004,Kim2009}. It has also  been observed that in certain observational learning tasks humans seem to follow a softmax reinforcement leaning scheme~\cite{Burke2010}. Thus, understanding softmax learning dynamics and its possible spectrum of behaviors is important both conceptually and for making concrete prediction about different learning outcomes. 

Here we use analytical techniques to provide a complete characterization of Boltzmann $Q$--Learning in two--player  two--action games, in terms of their convergence properties and rest point structure. In particular, it is shown that for any finite (non--zero) exploration rate, the learning dynamics necessarily converges to an interior rest point. This seems to be in contrast with previous observation ~\cite{Tuyls2006}, where we believe the authors have confused slow convergence with limit cycles. Furthermore, none of the studies so far have systematically examined the impact of exploration, i.e., {\em noise}, on the learning dynamics and its asymptotic behavior. On the other hand, noise is believed to be an inherent aspect of learning in humans and animals, either due to softmax selection mechanisms~\cite{Hopkins2002}, or random perturbations in agent utilities~\cite{Hofbauer2005}. Here we provide such an analysis, and show that depending on the game, there can be one, two, or three rest points, with a  bifurcation between different rest--point structures as one varies the exploration rate. In particular, there is a critical exploration rate above which there remains only one rest point, which is globally stable. 

The rest of this paper is organized as follows: We next describe the connection between Boltzmann $Q$-learning and replicator dynamics, and elaborate on the non--conservative nature of dynamics for any finite exploration rate. In Section~\ref{sec:analysis} we analyze the asymptotic behavior of the learning dynamics as a function of exploration rates for different game types. In Section~\ref{sec:examples} we illustrate our findings on several examples. We provide some concluding remarks in Section~\ref{sec:conclude}.

\section{Dynamics of Q--Learning}
\label{sec:QL}
Here we provide a brief review of $Q$-learning algorithm and  its connection  with the replicator dynamics. 

\subsection{Single Agent Learning}
In Reinforcement Learning (RL)~\cite{Sutton2000}  agents learn to behave near--optimally  through repeated interactions with the
environment.  At each step of interaction with the environment, the agent chooses an action based on the current state of the environment, 
and receives a scalar reinforcement signal, or a reward, for that action. The agent's overall goal is to learn to act in a way that  will increase  the long--term cumulative reward. 

Among many different implementation of the above adaptation mechanisms, here we consider the so called $Q$--learning~\cite{Watkins1992}, where the agents' strategies are parameterized through  $Q$--functions that characterize relative utility of a particular action. Those $Q$--functions are updated during the course of the agent's interaction with the environments, so that actions that yield high rewards are reinforced. To be more specific, assume that the agent has a finite number  of available actions, $i=1,2,...,n$, and let $Q_i(t)$ denote the $Q$-value of the corresponding action at time $t$. Then, after selecting action $i$ at time $t$, the corresponding $Q$-value is updated according to
\BEQ
Q_i(t+1) = Q_i(t) + \a [r_i(t) -  Q_i(t)]
\label{eq:update1}
\EEQ
where $r_i(t)$ is the observed reward for action $i$ at time $t$, and $\a$ is the learning rate. 

Next, we need to specify how the agent selects actions. Greedy selection, when the action with the highest $Q$ value is selected, might generally lead to globally suboptimal solution. Thus, one needs to incorporate some way of exploring less--optimal strategies. Here we focus on Boltzmann action selection mechanism, where the probability $x_i$ of selecting the action $i$ is given by  
\BEA
x_i(t) = \frac{e^{Q_i(t)/T}}{\sum_{k}e^{Q_k(t)/T}} \ , \ i=1,2,\cdots, n.
\label{eq:Boltzmann}
\EEA
where the {\em temperature} $T>0$ controls exploration/exploitation tradeoff: for $T\rightarrow  0$ the agent always acts greedily and chooses the strategy corresponding to the maximum $Q$--value (pure exploitation), whereas for $T\rightarrow \infty$ the agent's strategy is completely random (pure exploration).

We are interested in the continuous time limit of the above learning scheme. Toward this end, we divide the time into intervals $\delta t$, replace $t+1$ with $t+\delta t$ and $\a$ with $\a \delta t$. Next, we assume that within each interval $\delta t$, the agent samples his actions, calculates the average reward $r_i$ for action $i$, and applies Eq.~\ref{eq:update1} at the end of each interval to update the $Q$-values.\footnote{In the terminology  of reinforcement learning, this corresponds to an {\em off-policy} learning, as opposed to {\em on-policy} learning, where one uses Eq.~\ref{eq:Boltzmann} and Eq.~\ref{eq:update1} concurrently to sample actions and  update the $Q$-values of those action, respectively (e.g., see~\cite{Sutton2000}). A potential issue with the latter scheme is  that actions that are played rarely will be updated rarely, which might be problematics for the convergence of the algorithm. A possible remedy is to normalize each update amount by the frequency of corresponding action~\cite{Leslie2005,Sutton2000}, which can be shown to lead to the same dynamics Eq.~\ref{eq:Q2} in the continuous time limit.}

In the continuous time limit $\delta t \rightarrow 0$, one obtains the following differential equation describing the evolution of the $Q$ values: 
\BEQ
\dot{Q}_i (t)=  \a [r_i(t) -  Q_i(t)  ] 
\label{eq:Q2}
\EEQ

Next, we would like to express the dynamics in terms of strategies rather than the $Q$ values. Toward this end, we differentiate Eq.~\ref{eq:Boltzmann} with respect to time and use Eq.~\ref{eq:Q2}.  After rescaling the time, $t\rightarrow \a t/T$ , we arrive at the following set of equations:
\BEQ
 \frac{\dot{x_i}}{x_i} =    [r_i-\ssum_{k=1}^n x_k r_k] - T \ssum_{k=1}^n x_k\ln\frac{x_i}{x_k}.
 \label{eq:repx0}
\EEQ
The first term in Eq.~\ref{eq:repx0} asserts that the probability of taking action $i$ increases with a rate proportional to the overall efficiency of that strategy, while the second term describes the agent's tendency to {\em randomize} over possible actions. The steady state strategy profile, $x_i^s$, if it exists, can be found from equating the right hand side to zero, which can be shown to yield
\BEQ
\label{steady}
 x_i^s =    \frac{e^{r_i/T}}{\ssum_{k} e^{ r_k/T}}.
\EEQ
We would like to emphasize that  $x_i^s$ corresponds to the so called Gibbs distribution for a statistical--mechanical system with energy $-r_i$ at temperature $T$. Indeed, it can be shown that the above replicator dynamics minimizes the following  function resembling {\em free energy}:
\begin{equation}
\Phi[{\bf x}] =-\ssum_k r_kx_k + T\ssum_k x_k\ln x_k
\label{eq:functional}
\end{equation} 
where we have denoted ${\bf{x}}=(x_1,\cdots,x_n)$, $\ssum_{i=1}^nx_i=1$. Note that the minimizing the first term is equivalent to maximizing the expected reward, whereas minimizing the second term means maximizing the entropy of the agent strategy. The relative importance of those terms is regulated by the choice of the temperature $T$. We note that recently a free energy minimization principle has been suggested as a  framework for  modeling perception and learning (see~\cite{Friston2010} for a  review of the approach and its relation to several other neurobiological theories).

\subsection{Two-agent learning}
 Let us now  assume there are two agents that are learning concurrently, so that the rewards  received by the agents depend on their joint action. The generalization to this case is introduced via game-theoretical ideas \cite{Hofbauer2003}. More specifically, let $A$ and $B$ be the two payoff matrices: $a_{ij}$ ($b_{ij}$) is the reward of the first (second) agent when he selects $i$ and the second (first) agent selects $j$. Furthermore, let  ${\bf y}=(y_1,\cdots,y_n)$, $\ssum_{i=1}^n y_i=1$, be the strategy of the second agent. The expected rewards of the agents for selecting action $i$ are as follows:
  \BEA
 r_i^x = \ssum_{j=1}^n a_{ij}y_{j}  \ , \ r_i^y = \ssum_{j=1}^n b_{ij}x_{j} 
 \EEA
The learning dynamics in two-agent scenario case is obtained from Eq.~\ref{eq:repx0} by replacing $r_i$  with $r_i^x$ and $r_i^y$ for the first and second agents, respectively, which yields
 \BEA
  \dot{x}_{i}={x}_{i}[(A {\bf y})_{i}-{{\bf x}\cdot A{\bf y}}+T_X\sum_j x_j \ln( x_j/x_i )] \label{eq:repa} \\
 \dot{y}_{i}={y}_{i}[(B {\bf x})_{i}-{{\bf y}\cdot B{\bf x}}+T_Y\sum_j y_j \ln( y_j/y_i )] 
 \label{eq:repb}
 \EEA
 where $(A {\bf y})_{i}$ is the $i$ element of the vector $A\bf{y}$, and we assume that the exploration rates $T_X$ and $T_Y$ of the agents can generally be different. This system (without the exploration term) is known as bi--matrix replicator equation~\cite{Hofbauer1996,Hofbauer2003}. Its relation to multi--agent learning has been examined in~\cite{Borgers1997,Sato2002,Sato2003,Tuyls2003,Galla2009}. 
  
 Before proceeding further, we elaborate on the connection between the rest-points of the replicator system Eqs.~\ref{eq:repa},  \ref{eq:repb}, and the game-theoretic notion of Nash Equilibrium (NE), which is a central concept in game theory. Recall that a joint strategy profile $({\bf x}^*, {\bf y}^*)$ is called NE  if no agent can increase his expected reward by {\em unilaterally}  deviating from the equilibrium. It is known that for $T_X=T_Y=0$,  all the NE of a game are also rest-points of the dynamics~\cite{Hofbauer2003}. The opposite is not true -- not all the rest points correspond to NE. Furthermore,  some NE might correspond to unstable rest points  of the dynamics, which means that they cannot be achieved by the learning process. For any finite $T_X, T_Y>0$, the rest points will be generally different from the NE of the game. In the limit $T_X , T_Y \rightarrow \infty$, agents are insensitive to the rewards and mix uniformly over the actions. In this work we study the behavior of the learning dynamics in the intermediate  range of exploration rates.

\subsection{Exploration causes dissipation}
\label{dissipation}
It is known that for $T_X=T_Y=0$ the system of Eqs.~\ref{eq:repa},~\ref{eq:repb} are conservative~~\cite{Hofbauer1996,Hofbauer2003}, so that the total phase space volume is preserved. It can be shown, however, that any finite exploration rate $T_X,T_Y>0$ makes the system dissipative or volume contracting~\cite{Sato2003}. While this fact might not be crucial in high--dimensional dynamical system, its implications for low--dimensional system, and specifically for two--dimensional dynamical system considered here are crucial. Namely, the finite dissipation rate means that the system cannot have any limit cycles, and  the only possible asymptotic behavior is a convergence to a rest point. Furthermore, in situation when there is only one  interior rest point, it is guaranteed to be globally stable.

To demonstrate the dissipative nature of the system for $T_X,T_Y>0$, it is useful to make the following transformation of variables
\BEA
u_{k}=\ln \frac{x_{k+1}}{x_1} \ , \ v_{k}=\ln \frac{y_{k+1}}{y_1} \ , \ k=1,2,\cdots,n-1.
\EEA
The replicator system in the modified variables reads~\cite{Hofbauer1996,Sato2003}
\BEA
\dot{u}_{k} = \frac{\ssum_j {\tilde a}_{kj} e^{v_j}}{1+\ssum_j e^{v_j}} - T_X u_k \label{eq:u} \ , \
\dot{v}_{k} = \frac{\ssum_j {\tilde b}_{kj} e^{u_j}}{1+\ssum_j e^{u_j}} - T_Y v_k 
\EEA
where 
\BEA
{\tilde a}_{kj} = a_{k+1,j+1}-a_{1,j+1} \ , \
{\tilde b}_{kj} = b_{k+1,j+1}-a_{1,j+1}
\EEA
Let us  recall the Liouville  formula: If $ \dot{\bf z}={\bf F}({\bf z})$ is  defined on the open set U in $\mathbb{R}^n$ and if $G\subset U$ has volume $V(t)$ of $G(t)=\left \{{x(t):{\bf {x}} \in G }  \right \}$, then the rate of change of a volume V, which contain of set of points G  in the phase space is proportional to the divergence of {\bf F}~\cite{Hofbauer1998}.  Consulting with Eqs.~\ref{eq:u}, we observe that the dissipation rate is given by~\cite{Sato2003} 
\BEA
\ssum_k \biggl [\frac{\partial {{\dot u}_k}}{\partial u_k} + \frac{\partial {{\dot v}_k}}{\partial v_k} \biggr ] \equiv  -(T_X + T_Y)(n-1) <0
\EEA
As we mentioned above, the dissipative nature of the dynamics has important implications for two-action games that we consider next. 
 
 \subsection{ Two--action  games}
Let us consider two action games, and let $x$ and $y$ denote the probability of selecting the first action by the first and second agents, respectively. Then the learning dynamics Eqs.~\ref{eq:repa},~\ref{eq:repb} attain the following form: 
\BEA
\frac{\dot{x}}{x(1-x)}= \left (ay+b  \right)-\ln \frac{x}{1-x}\label{eq:repx}, \\ 
\frac{\dot{y}}{y(1-y)}= \left (cx+d  \right)-\ln \frac{y}{1-y} \label{eq:repy}
\EEA
where we have introduced
\BEA
a=-\frac{a_{21}+a_{12}-a_{11}-a_{22}}{T_X} \ , \  b=\frac{a_{12}-a_{22}}{T_X} 
\label{cof1}
\EEA
\BEA
c=-\frac{b_{21}+b_{12}-b_{11}-b_{22}}{T_Y} \ , \  d=\frac{b_{12}-b_{22}}{T_Y}
\label{cof2}
\EEA
The vertices of the simplex $\{x,y\}=\{0,1\}$ are rest points of the dynamics. For any $T_X,T_Y>0$, those rest points can be shown to be unstable. This means that any trajectory that starts in the interior of the simplex, $0<x,y<1$, will  asymptotically converge to an interior rest point. The position of those rest points is found by nullifying the RHS of Eqs.~\ref{eq:repx},~\ref{eq:repy}. For the remaining of this paper, we will examine the interior rest point equations in  details.

\section{Analysis of Interior Rest Points}
\label{sec:analysis}

\subsection{Symmetric  Equlibria}
\label{sec:sym}
\begin{figure}[!htb]
\centering
 \includegraphics[width = 0.4\textwidth]{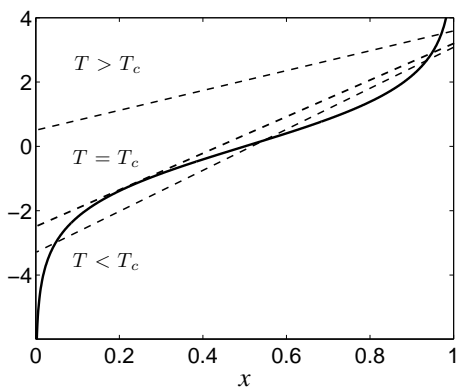}
  \caption{The graphical illustration of the rest point equation for the symmetric case,  Eq.~\ref{eq:fpsym}. The solid curve corresponds to the RHS, and the three lines  correspond  to the LHS for subcritical, critical and supercritical temperature values, respectively.} 
  \label{fig1}
\end{figure}

First, we consider the case of symmetric equilibria, $x=y$ and $T_X=T_Y=T$, in which case the interior rest point equation is 
\BEQ
ax+b  = \ln \frac{x}{1-x} 
\label{eq:fpsym}
\EEQ
Graphical representation of Eq.~\ref{eq:fpsym} is illustrated in Fig.~\ref{fig1} where we plot both sides of the equation as a function of $x$. First of all, note that the RHS of Eq.~\ref{eq:fpsym} is a monotonically increasing function, assuming values in $(-\infty, \infty)$ as $x$ changes between $(0,1)$. Thus, it is always guaranteed to have at least one solution. Further inspection shows that the number of possible rest points depends on the type of the game as well as the temperature $T$. 

For instance, there is a single solution whenever $a\le0$, for which  the LHS is a non--increasing function of $x$. 

Next, we examine the condition for having more than one rest point, which is possible when $a>0$. Consult with Fig.~\ref{fig1}: For sufficiently large temperature, there is only a single solution. When decreasing $T$, however, a second solution appears exactly at the point where the LHS becomes tangential to the RHS. Thus, in addition to Eq.~\ref{eq:fpsym}, at the critical temperature we should have 
\BEQ
a = \frac{1}{x(1-x)} \ , 
\label{eq:condsym}
\EEQ
or, alternatively,
\BEQ
x =   \frac{1}{2}\biggl [ 1  \pm \sqrt{1-\frac{4}{a}} \biggr ]
\label{eq:x}
\EEQ
Note  that the above solution exists only when $a\ge 4$. Plugging~\ref{eq:x} into  \ref{eq:fpsym}, we find 
 \BEQ
 b=\ln\frac{a\pm \a }{a\mp \a}-\frac{1}{2}(a\pm \a)  \ , \  \a=\sqrt{a^2-4a}
 \EEQ
Thus, for any given $a\ge 4$, the rest point equation has three solutions whenever $b_c^{-}<b<b_c^{+}$, where
 \BEA
b_c^{+}= \ln \frac{a-\a}{a+\a}-\frac{a-\a}{2} \ , b_c^{-} = \ln \frac{a+\a}{a-\a}-\frac{a+\a}{2}
\label{eq:b}
\EEA

\begin{figure}[!h]
\centering
    \includegraphics[width = 0.4\textwidth]{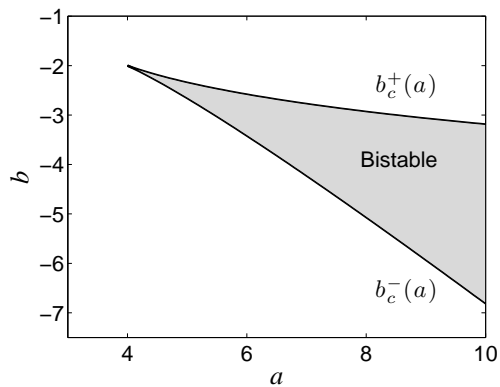}
       \caption{Demonstration of the cusp bifurcation in the space of parameters $a$ and $b$  for symmetric equilibria.}
        \label{fig2a}
\end{figure}

For small values of $T$ when $a$ is sufficiently large (and positive), the two branches $b_c^{-}$ and $b_c^{+}$ are well separated. When one increases $T$, however, at some critical value those two branches meet  and a cusp bifurcation occurs~\cite{Strogatz2001}. The point where the two bifurcation curves meet can be shown to be $(a,b)=(4,-2)$, and is   called a {\em cusp} point. A saddle-node bifurcation occurs all along the boundary of the region, except at the {\em cusp point}, where one has a codimension-2 bifurcation - i.e., two parameters have to be tuned for this type of bifurcation to take place~\cite{Strogatz2001}. This boundary in the parameter space is shown in Fig.~\ref{fig2a}.

\subsection{General Case}
\label{sec:gen}
\begin{figure}[!t]
\centering
    \includegraphics[width = 0.4\textwidth]{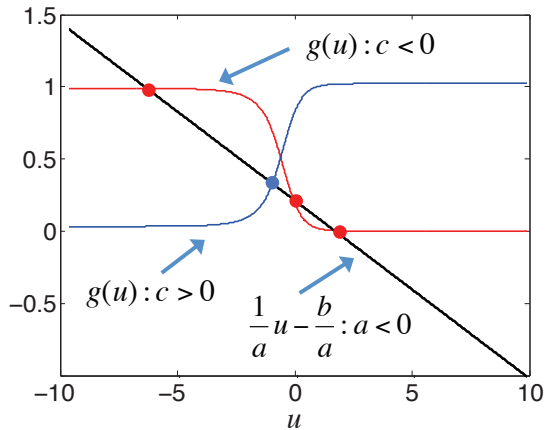} 
          \caption{(Color online) Graphical representation of the general rest point equation for two different values of $c$ : Intersections represent rest points. }
          \label{fig3a}  
\end{figure}

We now examine the most general case. We find it useful to introduce variables $u=\ln \frac{x}{1-x}$, $ v =\ln  \frac{y}{1-y}$. Then the interior rest point equations can be rewritten as   
\BEA
u = b + a \frac{1}{1+e^{-v}} \ , \ v = d + c \frac{1}{1+e^{-u}}
\label{eq:uvvariable}
\EEA
where $a$, $b$ , $c$,  and $d$ have been already defined  in Eqs.~\ref{cof1},~\ref{cof2}. Eliminating  $v$ we obtain
\BEA
\frac{1}{a}u - \frac{b}{a} =\biggl [1 + e^{-d-\frac{c}{1 + e^{-u}}} \biggr ]^{-1} \equiv g(u) \ .
\label{eq:rpnonsym}
\EEA
The solution of Eq.~\ref{eq:rpnonsym} are the rest point(s) of the dynamic. Its graphical representation is shown in Fig.~\ref{fig3a}. 

 It is easy to see that  $0<g(u)<1$. Furthermore, we have from Eq.~\ref{eq:rpnonsym}
\BEQ
g^{\prime}(u)=cg(1-g)\frac{1}{4\cosh^2\frac{u}{2}}
\label{eq:gderiv}
\EEQ
Thus, $g(u)$ is a  monotonically increasing (decreasing) function whenever $c>0$ ($c<0$). 
 
Next, we classify the games according to the number of rest points they allow. Let us consider two cases:

$\imath)$  $ac<0$: Note that in  Eq.~\ref{eq:rpnonsym} the LHS is a monotonically increasing (decreasing) function for $a>0$ ($a<0$). As stated above, RHS is also a monotonically increasing (decreasing) function whenever $c>0$ ($c<0$). Consequently, whenever $a$ and $c$ have different signs, i.e. $ac<0$, one of the sides is a monotonically increasing function while the other is a monotonically decreasing; thus, there can be only one interior rest point, which, due to the dissipative nature of the dynamics,  is globally stable. An example of this class of game is  Matching Pennies that will be discussed in Section~\ref{sec:examples}.

$\imath \imath)${ $ac>0$: } In this case it is possible to have one, two or three interior rest points. For the sake of concreteness, we focus on $a>0$, $c>0$, so that both the LHS  and  RHS of  Eq.~\ref{eq:rpnonsym}  are  monotonically increasing functions. 

Recall, that at the critical point when the second solution appears, the LHS of Eq.~\ref{eq:rpnonsym} should be tangential to $g(u)$. Consider now the set of  all tangential lines  to  $g(u)$ in Eq.~\ref{eq:rpnonsym}, and let $\delta_{min}$ and $\delta_{max}$ be the minimum and maximum value of the intercepts among  those tangential lines for any $u$ and $T_Y$. The intercept of the line given by the LHS of Eq.~\ref{eq:rpnonsym}, on the other hand, equals $-\frac{b}{a}$, and is independent of the temperature. It is straightforward to check that  multiple rest points are possible only when $\delta_{min}<-\frac{b}{a}<\delta_{max}$.

\begin{figure}[!h]
\centering
\includegraphics[width=0.40\textwidth]{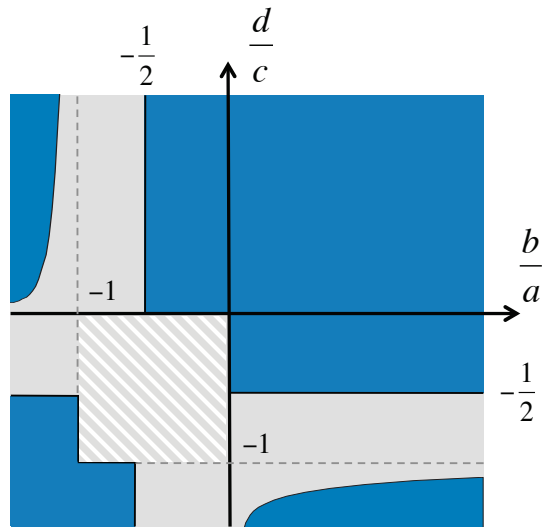}
\caption{(Color online) Characterization of different games in the parameter space with $a,c>0$. Dark blue region corresponds to games that can have only a single rest point, whereas the games in the light grey regions can have three rest-points. The shaded grey square corresponds to games that have three Nash equilibria.}  
\label{general}
\end {figure}

A full analysis along those lines  (see Appendix~\ref{sec:app1}) reveals that the number of possible rest points depend on the ratios $\frac{b}{a}$ and $\frac{d}{c}$, as depicted in Fig.~\ref{general}. First, consider the parameter range $0<-\frac{b}{a},-\frac{d}{c}<1$  (shaded light-grey region in Fig.~\ref{general}), which correspond to  so called coordination games that have three NE. The learning dynamics in these games can have three rest points, that intuitively correspond to the perturbed  NE. In particular, those rest points will converge to the NE as the exploration rates vanish. When $a,c<0$,  the parameter range $0<-\frac{b}{a},-\frac{d}{c}<1$ corresponds to so called anti-coordination games. Those games also have three NE, so the learning dynamics can have three rest points.

Let us now focus on  light grey (not-shaded) regions in Fig.~\ref{general}. The games in this parameter range have   a single NE. At the same time, the learning dynamics might still have multiple rest points. Those additional rest-points exist only for a range of exploration rates, and disappear when both exploration rates  $T_X, T_Y$ are sufficiently low or sufficiently high; see  Appendix~\ref{sec:app2} for details. An example of this type game will be presented  in Section~\ref{sec:examples}.

Note that the Fig.~\ref{general}  was obtained by assuming that $T_X$ and $T_Y$ are independent parameters. Assuming some type of functional dependence between those two parameters alters the above characterization. For instance, consider  the case $T_X=T_Y=T $. At the critical point we have (in addition to Eq.~\ref{eq:rpnonsym})   $ag^{\prime}(u)=1$, which yields
\begin{equation}
ac=\frac{4\cosh^2\frac{u}{2}}{g(1-g)}
\label{eq:cond2}
\end{equation} 
It can be shown\footnote{Indeed,  substituting $g(u)$ from Eq.~\ref{eq:rpnonsym} into Eq.~\ref{eq:cond2} one formally obtains a quadratic equation for $T$, $AT^2+BT+C=0$, $A=\frac{\cosh^2(u/2)}{{a}'{c}'}+u^2$, $B=(1+2\frac{{b}'}{{a}'})\frac{u}{{a}'}$, $C=(1+\frac{{b}'}{{a}'})\frac{{b}'}{{a}'}$ where: ${a}'={a_{21}+a_{12}-a_{11}-a_{22}}$ and ${c}'={b_{21}+b_{12}-b_{11}-b_{22}}$, ${b}'=a_{12}-a_{22}$. Requiring that $T$ is a real positive number yields $4AC<0$, or $0<-b/a<1$. With the similar reasoning the domain of $d/c$ of multiple intersection is $0<-d/c<1$.}  that when $T_X=T_Y=T $ the above conditions can be met only when $0<-\frac{b}{a}<1$, $0<-\frac{d}{c}<1$ (shaded region in Fig.~\ref{general}), which   correspond to the domain of  multiple NE: coordination  ($a,c>0$) and anti--coordination ($a,c<0$) games.  


It is illustrative to write  Eq.~\ref{eq:cond2} in terms of the original variables $x$ and $y$:
\begin{equation}
ac=\frac{1}{x(1-x)y(1-y)}
\label{eq:cond2b}
\end{equation} 
It can be seen that Eq.~\ref{eq:condsym} is recovered when $a=c$ and $x=y$. Furthermore, since $0<x,y< 1$, the above condition can be satisfied only when $ac\ge 16$.

\paragraph{Linear Stability Analysis}
We conclude this section by briefly elaborating on the dynamic stability of the interior rest points. Note that, whenever there is a single rest point  it will be globally stable due to the dissipative nature of the dynamics. Thus, we focus on the case when there are multiple rest points. 

For the interior rest points, the eigenvalues of the Jacobian of the dynamical system Eqs.~\ref{eq:repx},\ref{eq:repy}  are as follows:
 \begin{figure}[!t]
\centering   
    \subfigure[]{
    \includegraphics[width = 0.3\textwidth]{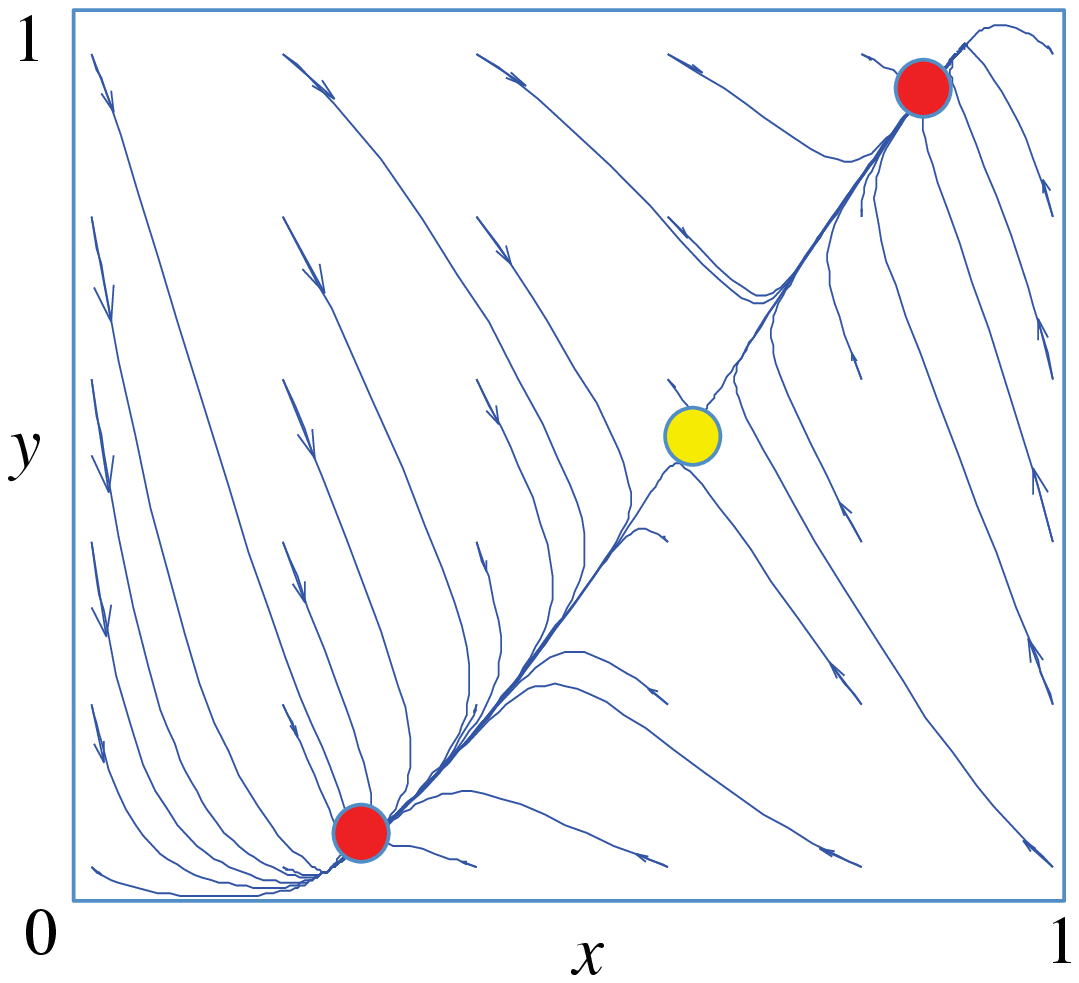} \label{fig5a}
    }
    \subfigure[]{
    \includegraphics[width = 0.3\textwidth]{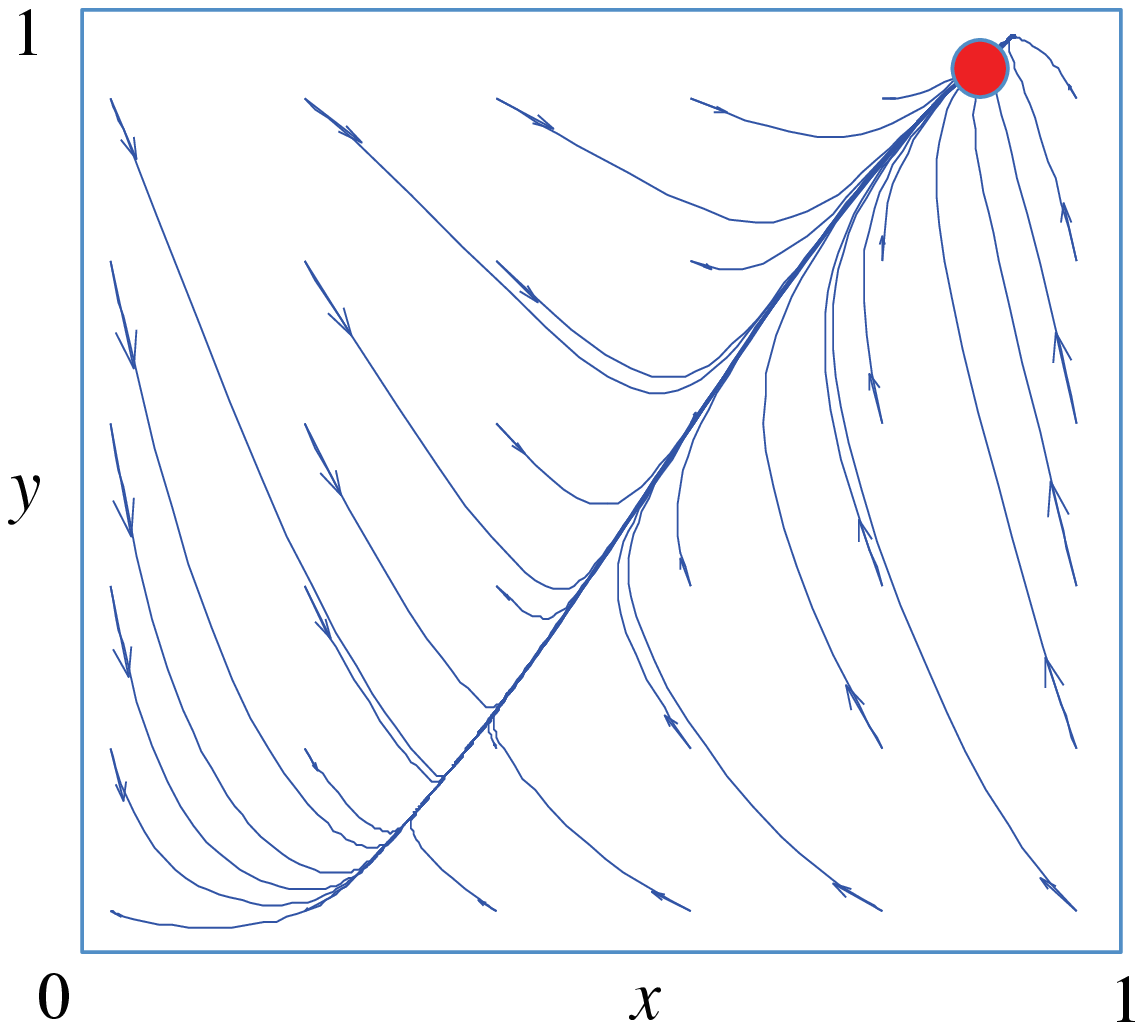} \label{fig5b}  
    }     
\caption{(Color online)  Illustration of dynamical flow for a system with three (a) and single (b) rest points.  Note that the middle rest point in (a) is unstable.}
\end{figure}
\begin{equation}
\lambda_{1,2}=-1\pm \sqrt{acy(1-y)x(1-x)}
\label{eigen}
\end{equation}
Let us focus on symmetric games and symmetric equilibria (i.e. $x=y$). From Eq.~\ref{eigen} we find the eigenvalues $\lambda_{1,2}=-1\pm {ax_0(1-x_0)}$, so that  the stability condition is $ax_0(1-x_0)<1$. Recalling that at the critical point we have $a=\frac{1}{x_0(1-x_0)}$, it is straightforward to demonstrate that for  the middle rest--point the above condition is always violated, meaning that it is always unstable. Similar reasoning shows that two other rest points are locally stable, and depending on the starting point of the learning trajectory, the system will converge to one of the two points. An example of the flows generated by the dynamics for below--critical and above--critical exploration rates is depicted in Fig.~\ref{fig5a} and~\ref{fig5b}. 

\section{Examples}
\label{sec:examples}
\begin{figure}[!h]
\centering
\includegraphics[width=0.45\textwidth]{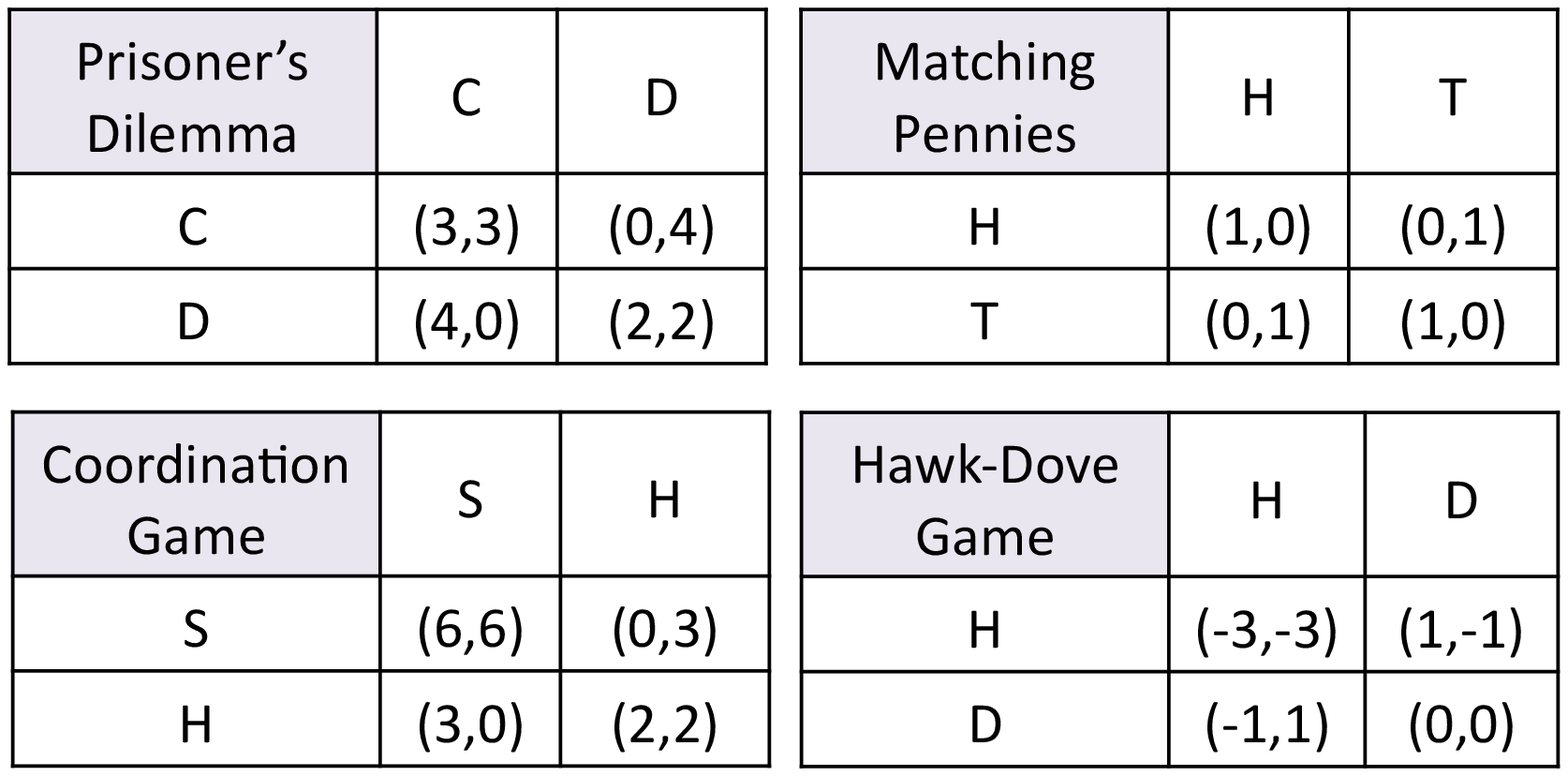}
\caption{Examples of reward matrices for  typical two-action games. }  
\label{figGames}
\end {figure}

We now illustrate the above findings on  several games shown in Fig.~\ref{figGames}. The row (column) number corresponds to the actions of the first (second) agent. Each cell contains a reward pair $(a_{ij},b_{ji})$, where  $a_{ij}$ and $b_{ji}$ are the corresponding elements of the reward matrices $A$ and $B$. 

 Our first example is the Prisoner's Dilemma (PD) where each player should decide whether to  {\em Cooperate} (C) or {\em Defect} (D). \comment{PD is characterized by parameter range $-$.} An example of a PD payoff matrix is shown in Fig.~\ref{figGames}. In PD the defection is a {\em dominant} strategy -- it always  yields a better reward regardless of the other player choice. Thus, even though it is beneficial for the players to cooperate, the only Nash equilibrium of the game is when both players defect. For $T_{X}=T_{Y}=0$, the dynamics always converges to the NE.

In our  PD example we have  $\frac{b}{a}=\frac{d}{c}=-2$, so according to Fig.~\ref{general}  there is a single interior rest point for any $T_{X},T_{Y}>0$. Furthermore, due to the dissipative nature of the dynamics, the system is guaranteed to converge to this rest point for any finite exploration rates. Note that this is in stark contrast from the behavior of $\epsilon$-greedy learning reported in~\cite{Wunder2010}, where the authors observed that, starting from some initial conditions, the dynamics  might never converge, instead alternating between different strategy regimes. The lack of convergence and chaotic behavior in their case can be attributed to the hybrid nature of the dynamics. 

Next, we consider Matching Pennies (MP), which is a zero sum game where the first (second) player wins if both players select the same (different) actions; see Fig.~\ref{figGames}. This game does not have any pure NE, but it  has a mixed NE at $x^*=y^*=\frac{1}{2}$. This mixed NE is a rest point of the learning dynamics at $T_{X}=T_{Y}=0$ which is a $center$ point surrounded by periodic orbits~\cite{Hofbauer1996}. For this game we have  ${a}{c}<0$. Thus, there can be only one interior rest point, which can be globally stable for any $T_{X},T_{Y}>0$. Furthermore, a particular feature of this game is that finite $T_{X}, T_{Y}$ does not perturb the position of the rest-point (since the entropic term is zero for $x=y=\frac{1}{2}$). 

We now consider  a coordination game (shaded area in Fig.~\ref{general}) where players have an incentive to select the same action. In the example shown in Fig.~\ref{figGames}, the players should decide whether to hunt a {\em stag} (S) or a {\em hare} (H). This game has two pure NE,  (S,S) and (H,H),  as well as  a mixed NE at $(x^*,y^*)=(-\frac{b}{a},-\frac{d}{c})$, which, for the particular coordination game shown in Fig.~\ref{figGames}, yields  $x^*=y^*= 2/5$. For sufficiently small exploration rates, the learning dynamics has three rest points that intuitively correspond to  the three NE of the game. Furthermore,  the rest points corresponding to the pure equilibria are stable, while the one corresponding to the mixed equilibrium is unstable.

\begin{figure}[!t]
\centering
       \subfigure[]{
    \includegraphics[width = 0.315\textwidth]{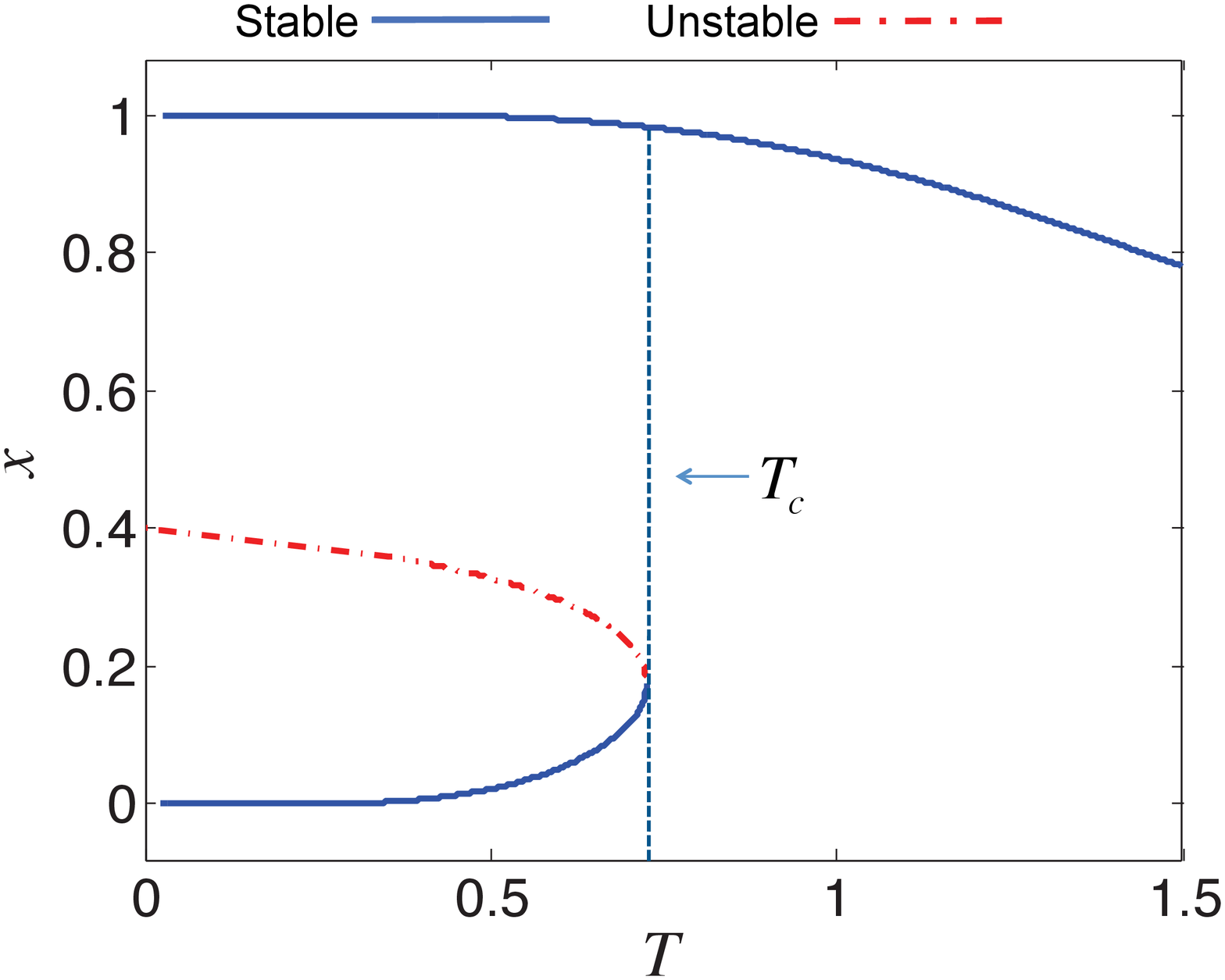} \label{fig7a}  
    }   
     \subfigure[]{
    \includegraphics[width = 0.315\textwidth]{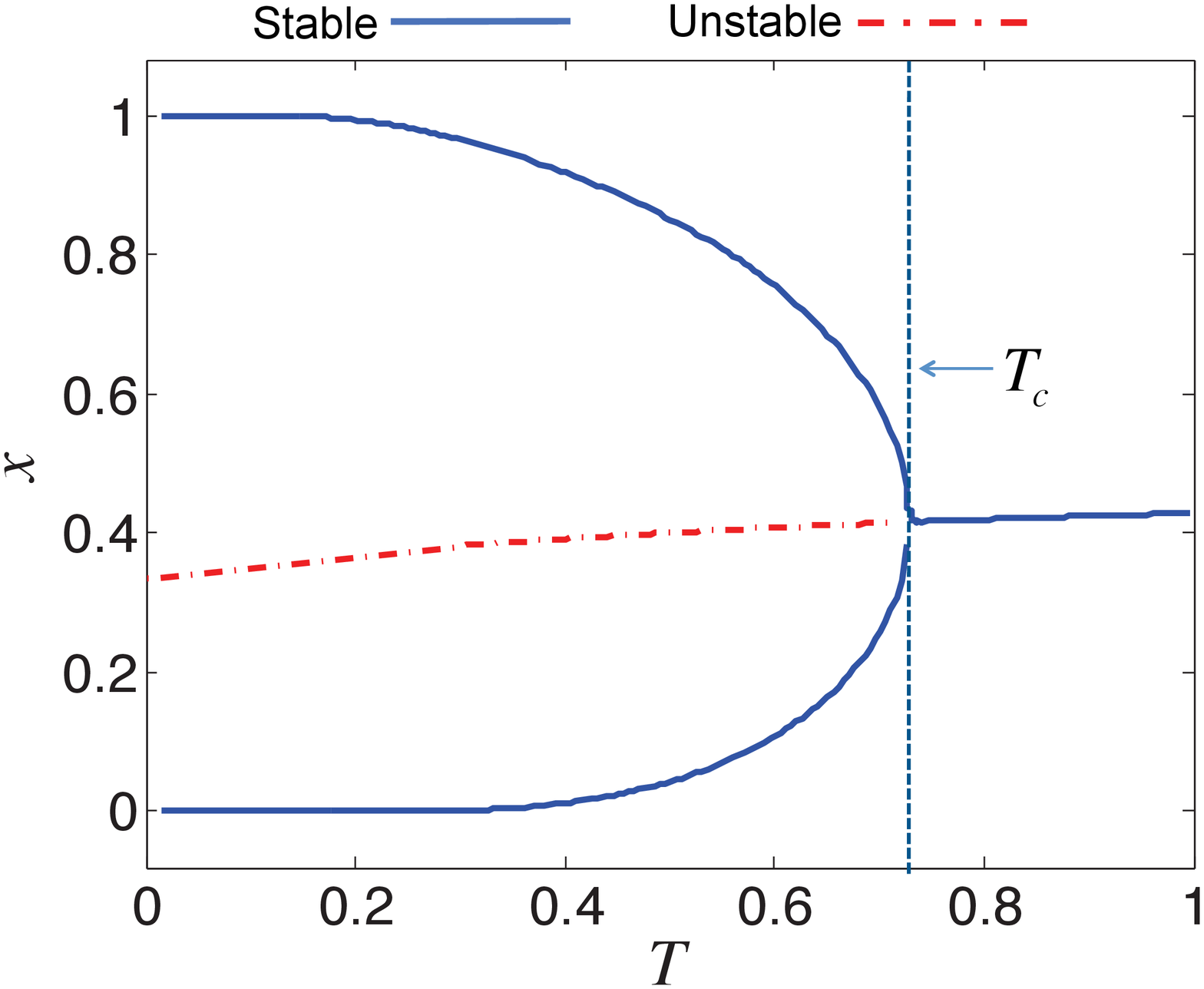} \label{fig7b}
    }
       \caption{(Color online)  Bifurcation  diagram of the rest points for $T_X=T_Y=T$ : (a) Disconnected pitchfork, with mixed NE: $(x^*,y^*)=(2/5, 2/5)$ (b) Continuous pitchfork, with mixed NE: $(x^*,y^*)= (1/3, 2/3)$.}
        \label{SymBif}
\end{figure}

When increasing the exploration rates, there is a critical line $(T_X^c, T_Y^c)$ so that for any $T_X>T_X^c, T_Y>T_Y^c$ only one of the rest points  survives. In Fig.~\ref{SymBif} we show   the bifurcation diagram on the plane $T_X=T_Y$.\footnote{We find that the bifurcation structure is qualitatively similar for the more general case  $T_X \neq T_Y$.} We find that most coordination games are characterized by a discontinuous pitch-fork bifurcation (see Fig.~\ref{fig7a}), where above the critical line the surviving rest point correspond to the {\em risk-dominant} NE~\footnote{In a general coordination game,  the strategy profile (1,1) is risk dominant if $(a_{12}-a_{22})(b_{12}-b_{22})\ge(a_{21}-a_{11})(b_{21}-b_{11})$. In  symmetric coordination games (i.e., as shown in Fig.~\ref{general}) the strategy profile is  risk-dominant if it yields a better payoff against an opponent that plays a uniformly mixed strategy.}. There is an exception, however, for  games  with $\frac{b}{a}+\frac{d}{c}=-1$. This condition describes games where none of the pure NE are strictly risk dominant, and where the mixed NE  satisfies $x^*+y^*=1$.  The rest point structure undergoes a continuous pitchfork bifurcation as shown Fig.~\ref{fig7b} whenever  $a=c$ and $\frac{b}{a}+\frac{d}{c}=-1$. One can show that when the above condition is met, the critical point  $u_0$ that satisfies $g^{\prime}(u_0)=\frac{1}{a}$ , $\frac{1}{a}u_0 - \frac{b}{a}=g(u_0)$, is also the inflection point of $g(u)$, $g^{\prime \prime}(u_0)=0$.

The other class of two-action games with multiple NE are so-called anti--coordination  games where it is mutually beneficial for the players to select different actions. In anti--coordination games, one has $a,c<0$ whereas  $0<-\frac{b}{a}<1$, $0<-\frac{d}{c}<1$. A popular example  is the so called Hawk-Dove game where players should choose between an aggressive (H) or peaceful (D) behavior. This game has two pure NE,  (H,D), (D,H), and a mixed NE at $(x^*,y^*)=(-\frac{b}{a},-\frac{d}{c})$. An example is shown in Fig.~\ref{figGames} with a mixed NE at $x^*=y^*= 1/3$.

Anti-coordination games have similar bifurcation structure compared to the coordination games. Namely,  there is a critical line $(T_X^c, T_Y^c)$ so that for any $T_X>T_X^c,T_Y>T_Y^c$ only a single rest point survives. As in the coordination games, the bifurcation is discontinuous for most parameter values. The condition for continuous pitch-fork bifurcation in the anti-coordination games is given by $a=c$ and $\frac{b}{a}=\frac{d}{c}$. Thus, those games have a symmetric NE $x^{*}=y^{*}$. Furthrmore, the critical point where the second solution appears is also the inflection point of $g(u)$, $g^{\prime \prime}(u_0)=0$. 

\begin{figure}[!t]
\centering
       \subfigure[]{
    \includegraphics[width = 0.315\textwidth]{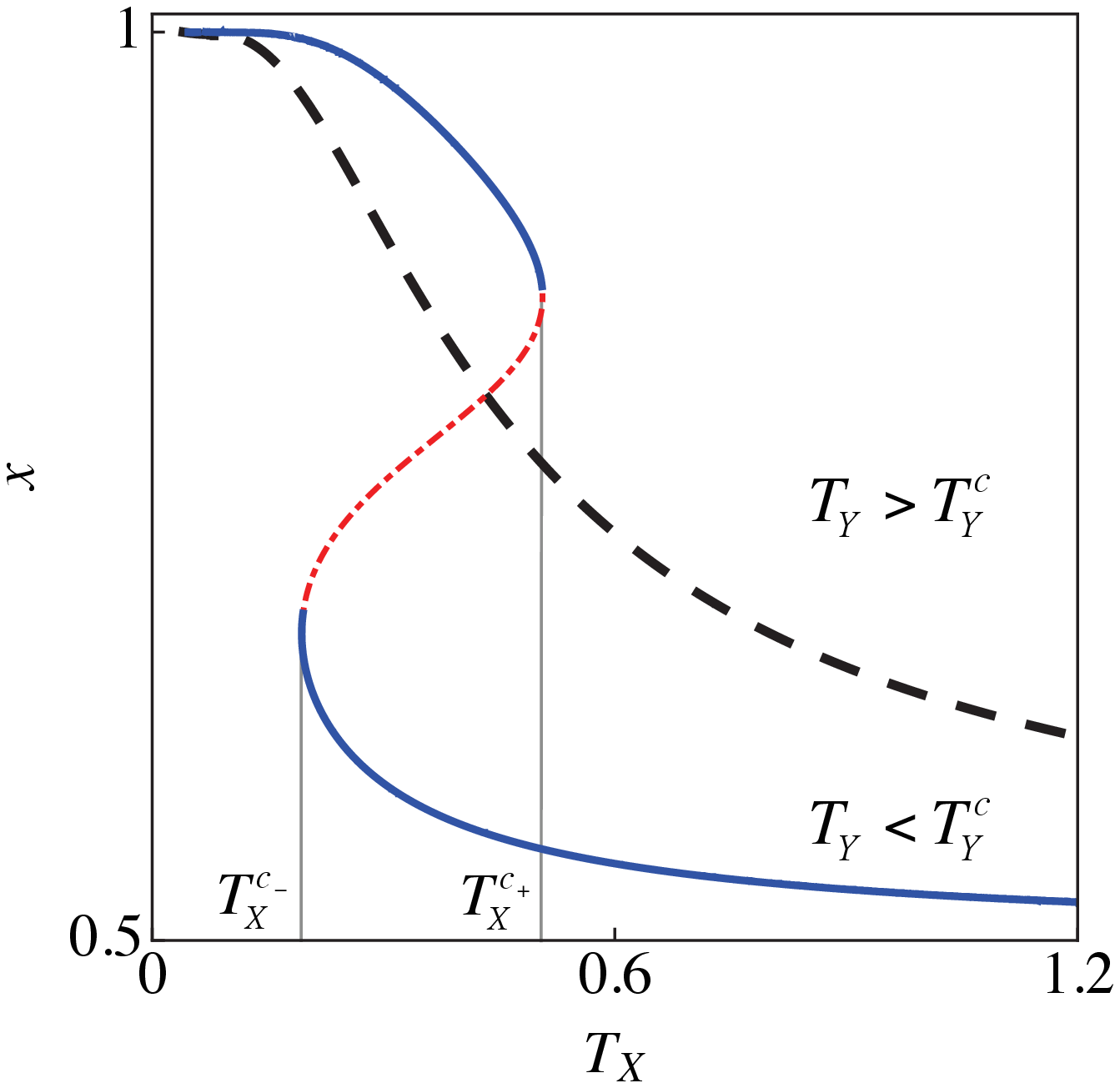} \label{fig8a} 
    }  
     \subfigure[]{
    \includegraphics[width = 0.315\textwidth]{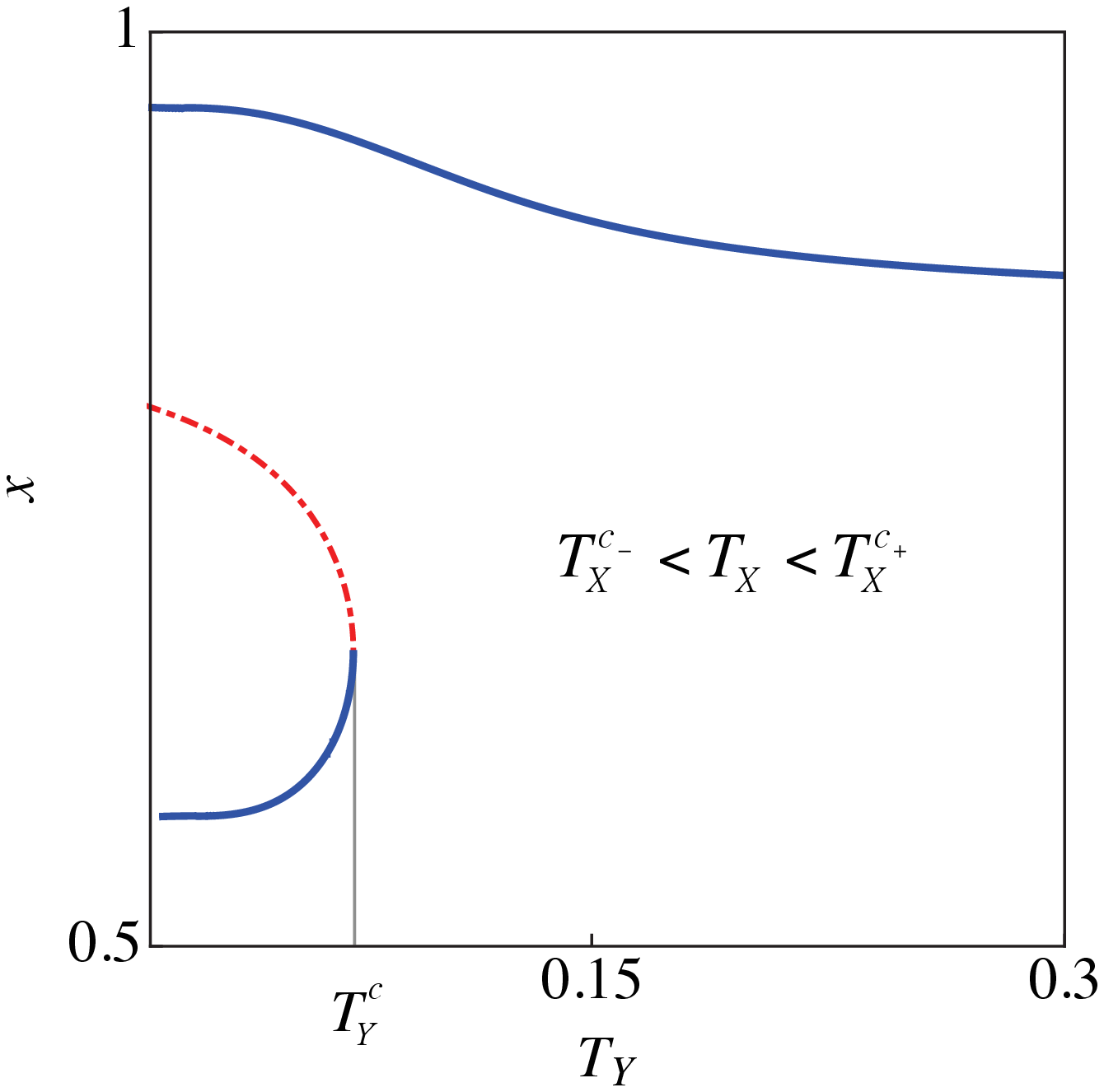} \label{fig8b}
    }
     \caption{(Color online) Bifurcation in  the domain  of the games with $a,c>0$, $\frac{b}{a}>0$, $ -\frac{1}{2}> \frac{d}{c}>-1$. In this example we have:
$\frac{d}{c} = -0.8, \frac{b}{a} = 0.1$:  a)  Rest point structure plotted against $T_X$ for $T_Y<T_Y^c$ and $T_Y>T_Y^c$. b) The rest point structure plotted against $T_Y$ for  $T_X^{c_{-}}<T_X<T_X^{c_{+}}(T_Y)$. In both graphs, the red dot-dashed lines correspond to  the unstable rest points.}
\label{Bifu}  
\end{figure}

Finally, let us consider the games with a single NE, for which the learning dynamics can still have multiple rest points. To be specific, we focus on the case $a,c>0$, for which the possible regimes are outlined in Fig.~\ref{general}. In Fig.~\ref{fig8a} , we show the dependence of the rest point structure on the parameter $T_X$, for two different values of $T_Y$, for $\frac{b}{a}=0.1$, $\frac{d}{c}=-0.8$. It can be seen that for sufficiently small $T_X$,  the learning dynamics  allows a single rest point (that corresponds to the NE of the game). Similarly, there is single rest points whenever  $T_Y$ is sufficiently hight.  However,  there is a critical exploration rate for agent $Y$, $T_Y^c$, so that for any $0<T_Y<T_{Y}^{c}$, there is a range  $T_X^{c_{-}}(T_Y)<T_X<T_X^{c_{+}}(T_Y)$, for which the dynamics allows three rest points. In contrast to coordination and anti--coordination games considered above, those additional rest points do not correspond  to any NE of the game. In particular, they disappear when $T_X, T_Y$ are sufficiently small. We elaborate more on the appearance of those rest points in Appendix~\ref{sec:app2}.

Fig.~\ref{fig8b} shows the bifurcation diagram for the same game but plotted against $T_Y$. Note that the two diagrams are asymmetric. In particular, in contrast to Fig.~\ref{fig8a}, here multiple solutions are possible even when $T_Y$ is arbitrarily small (provided that $T_X^{c_{-}}(T_Y)<T_X<T_X^{c_{+}}(T_Y)$).  This asymmetry is due to the fact that the agents' payoff matrices represent different games. In this particular case, the first player's payoff matrix corresponds to a dominant action game, whereas the second player's payoff matrix corresponds to a coordination game. Clearly, when $T_X$ is very small,  the first player  will mostly select the dominant action, so there can be only a single rest point at small $T_X$. Increasing $T_X$ will make the entropic term more important, until at a certain point, multiple rest points will emerge. 

The same picture is preserved for the parameter range $\frac{b}{a}<-1, -\frac{1}{2}<\frac{d}{c}<0$ (the other light grey horizontal stripe). On the other hand, the players effectively exchange roles in the parameter ranges corresponding to the vertical stripes: $\frac{d}{c}>0,-1<\frac{b}{a}<-\frac{1}{2}$ and  $\frac{d}{c}<-1,-\frac{1}{2}<\frac{b}{a}<0$. In this case,  there is a critical exploration rate $T_X^c$, so that for any $0<T_X<T_{X}^{c}$, there is a range  $T_Y^{c_{-}}(T_X)<T_Y<T_Y^{c_{+}}(T_X)$, for which the dynamics allows three rest points. 

Finally, we note that the rest point behavior is different in the light grey regions where the parameters are also confined to $\frac{b}{a}>0, \frac{d}{c}<-1$ and  $\frac{b}{a}<-1, \frac{d}{c}>0$. In those regions, multiple rest points are available only when both $T_X$ and $T_Y$ are strictly positive, i.e.,  $T_X^{c_{-}}>0$, $T_Y^{c_{-}}>0$.

\section{Discussion}
\label{sec:conclude}

We have  presented a comprehensive analysis of two agent $Q$--learning dynamics with Boltzmann action selection mechanism, where the agents exploration rates are governed by temperatures $T_X, T_Y$. For any two action game at finite exploration rate  the dynamics is dissipative and thus guaranteed to reach a rest point asymptotically. We demonstrated that, depending on the game and the exploration rates, the rest point structure of the learning dynamics is different. When $T_X=T_Y$, for games with a single NE (either pure or mixed) there is a single globally stable  rest point  for any positive exploration rate. Furthermore, we analytically examined the impact of exploration/noise on the asymptotic behavior, and showed that in games with multiple NE the rest--point structure undergoes a bifurcation so that above a critical exploration rate only one globally stable solution persists. Previously, a similar observation for certain games was observed numerically in Ref.~\cite{Wolpert2010}, where the authors studied  Quantal Response Equilibrium (QRE) among agents with bounded rationality. In fact, it can be shown that QRE corresponds to the rest--point of the  Boltzmann $Q$-learning dynamics. A similar bifurcation pictures was also demonstrated for certain continuous action games~\cite{Galstyan2011JAAMAS}. 

In  general, we observed that for $T_X\neq T_Y$, the learning dynamics  is  qualitatively similar for  games with multiple NE. Namely, there is a bifurcation at critical exploration rates $T_X^c$ and $T_Y^c$, so that the learning dynamics allows three (single) rest points below (above) those critical values. In particular, the rest points converge to the NE of the game when $T_X,T_Y \rightarrow 0$. What is perhaps more interesting is that for certain games with a single NE, it is possible to have multiple rest points in the learning dynamics when $T_X \neq T_Y$. Those additional rest points persist only for a finite range of exploration rates, and disappear when the exploration rates $T_X$ and $T_Y$ tend to zero.

We suggest that the sensitivity of the learning dynamics on exploration rate can be useful for validating various hypotheses about possible learning mechanisms in experiments. Indeed, most empirical studies so far have been limited to games with a single equilibrium, such as matching pennies, where the dynamics is rather insensitive to the exploration rate. We believe that for different games (such as coordination or anti-coordination game), the fine--grained nature of the rest point structure, and specifically, its sensitivity to the exploration rate, can provide much richer information about learning mechanisms employed by the agents. 

{\em Note Added:} After completing the manuscript, we became aware of a very recent work reporting similar results~\cite{Kaiser2011}, which studies convergence properties and bifurcation in the solution structure using local stability analysis. For games with a single rest point such a Prisoner's Dilemma, local stability is subsumed by the global stability demonstrated here. The bifurcation results are similar, even though \cite{Kaiser2011} studies only coordination games and does not differentiate between continuous and discontinuous pitchfork bifurcation. Finally,  the analytical form of the phase diagram Eq.~\ref{eq:b} for the symmetric case, as well as the possibility of multiple rest points for  games with a single NE demonstrated here, are complementary to the results presented in~\cite{Kaiser2011}.

 \section{ Acknowledgments}
We thank Greg Ver Steeg for useful discussions.  This research was supported in part by the National Science Foundation under grant No. 0916534 and the US AFOSR MURI grant No. FA9550-10-1-0569.


\appendix

\section{Classification of games according to the number of allowable rest-points}
\label{sec:app1}
Here we derive the conditions for multiple rest-points. We assume $a,c>0$ for the sake of concreteness. 

Consider the set of all the tangential lines to $g(u)$ (see Eq.~\ref{eq:rpnonsym}), and let $\delta_{T_Y}(u)$ be the intercept of the tangential line that passes through point $u$,  $\delta_{T_Y}(u)=g(u)-g^{\prime}(u)u$: Here the subscript indicates that the intercept depends on the exploration rate $T_Y$ via coefficients $c$ and $d$. 
 The extremum of function $\delta_{T_Y}(u)$ happens at $\frac{d \delta_{T_Y}}{du}=-g^{\prime \prime}u=0$ where:
\BEQ
g^{\prime \prime}(u)=-\frac{cg(1-g)}{16\cosh^4\frac{u}{2}} \biggl ( c \tanh  \biggl [ \frac{d}{2}  + \frac{c/2}{1+e^{-u}} \biggr ] +2 \sinh u \biggr )
\label{eq:gderive2}
\EEQ
Let $u_0$ be the point where $g^{\prime \prime}(u_0)=0$. A simple analysis yields that $u_0>0$ whenever $d<-c/2$, and $u_0<0$ otherwise. Next, let $\delta_{min}=\min_{u,T_Y}\delta_{T_Y}(u)$ and $\delta_{max}=\max_{u,T_Y}\delta_{T_Y}(u)$, where minimization and maximization is over both $u$ and $T_Y$. It can be shown that there can be  multiple solutions only when $\delta_{min}<-\frac{b}{a}<\delta_{max}$.

We now consider different possibilities depending on the ratio $\frac{d}{c}$. Due to symmetry, it is sufficient to  consider $\frac{d}{c}<-\frac{1}{2}$. We differentiate the following cases:

$\imath) -1<\frac{d}{c}<-\frac{1}{2}$: In this case one has $\delta_{min}=-\infty$, $\delta_{max}=1$. Thus, there will be one rest point whenever $\frac{b}{a}<-1$. 

$\imath \imath) \frac{d}{c}<-1$: In this case one has $\delta_{max}=\frac{1}{2}$, thus, there will be single rest point whenever $\frac{b}{a}<-\frac{1}{2}$. Furthermore, although an analytical  expression for $\delta_{min}$ is not available, the corresponding boundary can be found by numerically solving a transcendental equation $-\frac{b}{a}=\delta_{min}$ for different $\frac{d}{c}$.

Repeating the same reasoning for  $\frac{d}{c}>-\frac{1}{2}$ yields the different regions depicted in Fig.~\ref{general}. 

\vspace{10mm}

\section{Appearance of multiple rest points in games with single NE}
\label{sec:app2}
\begin{figure}[!t]
\centering
 \includegraphics[width = 0.4\textwidth]{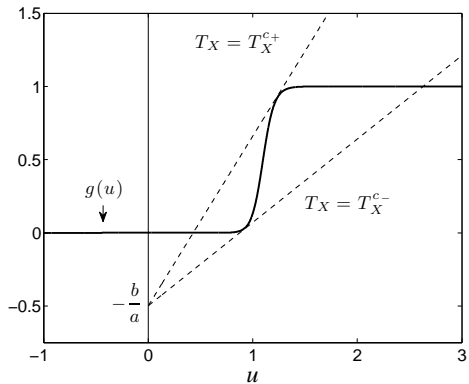}
\caption{Graphical illustration of the multi-rest point equation for a game with a single NE. Here $a,c>0$, $\frac{b}{a}=\frac{1}{2}$, $\frac{d}{c}=-\frac{3}{4}$. }  
\label{fig9}  
\end {figure}

We now elaborate on games with single NE for which the learning dynamics still can have multiple rest points. For the sake of concreteness, let us consider one of the regions in Fig.~\ref{general} that corresponds to $\frac{b}{a}>0$, $-1<\frac{d}{c}<-1/2$. The graphical representation of the rest point equation  is shown in Fig.~\ref{fig9}. For a given $T_Y$, the two lines correspond to the critical values of $T_X^{c_{-}}(T_y)$ and $T_X^{c_{+}}(T_Y)$. Let us consider the case $T_Y=0$. It is easy to see that in this limit $g(u)$ becomes a step function, $g(u)=\theta(u-{ \tilde u})$, where ${ \tilde u}$ is found by requiring $\frac{d}{c}+\frac{1}{1+e^{-u}}=0$, which yields $\tilde{u}=\ln \frac{-d}{d+c}$. Simple calculations yield $T_X^{c_{-}}(T_Y=0)=\frac{\tilde a}{\tilde u}\frac{b}{a}$ and  $T_X^{c_{-}}(T_Y=0)=\frac{\tilde a}{\tilde u}(\frac{b}{a}+1)$, where $\tilde a =a_{21}+a_{12}-a_{11}-a_{22}\equiv aT_X$. For general $T_Y>0$, the corresponding values  $T_X^{c_{-}}(T_Y)$ and $T_X^{c_{+}}(T_Y)$ can be found numerically. 
Finally, note that when increasing $T_Y$, there is a critical exploration rate $T_Y=T_Y^c$ so that for $T_Y>T_Y^c$ the multiple solutions will disappear. It is easy to see that $T_Y^c$ corresponds to the point when the maximum value of the intercept to $g(u)$ for a given $T_Y$ equals  $-\frac{b}{a}$.


\end{document}